\documentclass [12pt]{article}
\usepackage{latexsym}
\usepackage{amsfonts}
\usepackage{acronym}
\newcommand{\be}{\begin{equation}}
\newcommand{\ee}{\end{equation}}
\newcommand{\bea}{\begin{eqnarray}}
\newcommand{\eea}{\end{eqnarray}}
\makeatletter
\@addtoreset{equation}{section}

 \makeatother
\begin{document}
\normalsize
\title{\Large Superfluidity of  Minkowskian Higgs vacuum with BPS monopoles quantized by Dirac may be described as  Cauchy problem to Gribov ambiguity equation.}
\author
{{\bf L.~D.~Lantsman}\\
 Wissenschaftliche Gesellschaft bei
 J$\rm \ddot u$dische Gemeinde  zu Rostock,\\Augusten Strasse, 20,\\
 18055, Rostock, Germany; \\ 
Tel.  049-0381-799-07-24,\\
llantsman@freenet.de}
\maketitle
\begin {abstract}
We show that manifest superfluid properties of the Minkowskian Higgs model with vacuum BPS monopoles quantized by Dirac may be described in the framework of the Cauchy problem to the Gribov ambiguity equation.

The latter  equation specifies the ambiguity in choosing the covariant Coulomb (transverse) gauge for Yang-Mills fields represented as topological Dirac variables, may be treated as solutions to the Gauss law constraint at the removal of temporal components of these fields.

We demonstrate that the above Cauchy problem comes just to fixing the covariant Coulomb gauge for topological Dirac variables in the given initial time instant $t_0$ and finding the solutions to the Gribov ambiguity equation in the shape of vacuum BPS monopoles and excitations over the BPS monopole vacuum referring to the class of multipoles.

The next goal of the present study will be specifying  the look of Gribov topological multipliers entering Dirac variables in the Minkowskian Higgs model quantized by Dirac, especially at the spatial infinity, $\vert {\bf x}
\vert \to \infty$ (that corresponds to the infrared region of the momentum space).   
\end{abstract}
 \newpage
\tableofcontents
\newpage
\section{Introduction.}
In the recent paper \cite{rem1} the Minkowskian Higgs model with vacuum BPS monopole solutions \cite{Al.S.,BPS, Gold,LP2,LP1} was studied in the framework of the Faddeev-Popov (FP) "heuristic" quantization scheme  \cite{FP1}.

The essence of the FP "heuristic" quantization approach is in fixing a gauge: say, $F(A)=0$ in FP path integrals \cite{FP1}.

In particular, as it was demonstrated in \cite{rem1} (repeating the arguments \cite{Al.S.}), the FP "heuristic" quantization \cite{FP1} of the Minkowskian Higgs model with vacuum BPS monopoles may be reduced to fixing the {\it temporal} (Weyl) gauge $A_0=0$ for temporal Yang-Mills (YM) components via the $\delta(A_0)$ multiplier in the appropriate FP path integral.

For stationary BPS monopole solutions it is equivalent to vanishing "electric" fields $F_{0i}^a$ in the above Minkowskian Higgs model.

There may be demonstrated \cite{rem1} that the absence of "electric" fields $F_{0i}^a$ in the \linebreak Minkowskian Higgs model with vacuum BPS monopole solutions prevents any (topologically) nontrivial dynamics in that model.

The same proves to be correct also for another Minkowskian (Higgs) models with vacuum monopoles: for instance, in the 't Hooft-Polyakov model \cite{H-mon, Polyakov} or in the Wu-Yang one \cite{Wu}.

\medskip
An essential point of all the Minkowskian (Higgs) models with vacuum monopoles quantized in the FP "heuristic" \cite{FP1} wise is assuming the "continuous"
$$  SU(2)/ U(1)\sim S^2 \equiv R           $$ 
vacuum geometry therein.

This implies point hedgehog topological defects in these Minkowskian Higgs models.

As it was explained in Ref.  \cite{Al.S.}, the origin of point hedgehog topological defects in Minkowskian Higgs models with vacuum monopole solutions is in the isomorphism
$$ \pi_2 R= \pi_1 U(1) \equiv  \pi_1 H,     $$
taking place obviously for the residual $ U(1)$ gauge symmetry group inherent in these models.

\medskip
In the paper \cite{rem1} there was also argued that the Minkowskian Higgs model with vacuum BPS monopoles is  the unique Minkowskian model with monopoles in which the manifest superfluidity of the vacuum takes place.

It is induced by the {\it Bogomol'nyi equation} \cite{Al.S.,Gold,LP2,LP1}
\be \label{Bog}  {\bf B} =\pm D \Phi,      \ee
giving the relation between the vacuum "magnetic" field $\bf B$ and the Higgs isomultiplet $\Phi$ (taking the shape of a BPS monopole).

Herewith the transparent parallel between the BPS monopole vacuum (quantized in the FP "heuristic" \cite{FP1} wise) and the superfluid component in a liquid helium II specimen \cite {N.N.} was pointed out in \cite{rem1}.

This parallel comes to the treatment of the Bogomol'nyi equation as the {\it potentiality condition} for the BPS monopole vacuum.

In this case the vacuum "magnetic" field $\bf B$ plays the same role  that the (critical) velocity ${\bf v}_0$ \cite {N.N.} of the superfluid motion in a liquid helium II specimen.

\medskip
On the other hand, to the end  of the discussion in Ref. \cite {rem1},  there was noted that the Bogomol'nyi equation and manifest superfluid properties of the Minkowskian Higgs model with BPS monopoles (generated by this equation)  prove to be compatible with the Dirac fundamental quantization \cite {Dir} as well as with the FP "heuristic" one \cite {FP1} of that model. \par

A brief analysis of the Dirac fundamental quantization scheme \cite {Dir} conformably to the Minkowskian Higgs model with BPS monopole solutions was performed recently in the paper \cite {fund}.

This comes to the Gauss-shell reduction of the Minkowskian (YM-Higgs) action functional in terms of topological Dirac variables $\hat A^D_i$ ($i=1,2$) \cite {David2, David3, Pervush2}: transverse and gauge invariant functionals of YM fields.  

The transverse gauge for topological Dirac variables $\hat A^D_i$ may be written down as \cite {rem1, fund, David2}
\be \label{Dv} \partial_0 D_i {\hat A}^D_i({\bf x},t) =0.\ee
On the other hand, there is an ambiguity in choosing the  transverse gauge (\ref{Dv}) for topological Dirac variables $\hat A^D$: there are always two  YM fields $\hat A^D$ and $\hat A_1^D$ satisfying (\ref{Dv}). 

It is a purely non-Abelian effect discovered in the paper \cite {Gribov} by V. N. Gribov and called the {\it Gribov ambiguity}.
  The topological causes of the Gribov ambiguity were also enough good revealed in the monograph \cite {Al.S.} (in \S T26).

\medskip
On the level of the Dirac fundamental quantization \cite {Dir} of the Minkowskian Higgs model with BPS monopoles, the Gribov ambiguity in specifying the  transverse gauge (\ref{Dv}) for topological Dirac variables may be expressed \cite {rem1,LP2,LP1, Pervush1} with the aid of the {\it Gribov ambiguity equation}
\be \label{p. con}  D^2\Phi=0     \ee
imposed onto the Higgs field $\Phi$ having the look of a vacuum BPS monopole. \par 
This "potentiality condition" comes to the Bogomol'nyi  equation due to the Bianchi identity
$$ D B=0.   $$
Thus the Bogomol'nyi  equation, derived \cite{Al.S.,BPS} in the temporal gauge $A_0=0$ for YM fields (and manifest superfluid properties of the Minkowskian BPS monopole vacuum induced by this equation), proves to be compatible with the FP "heuristic" \cite{FP1} as well as with the Dirac "fundamental" quantization \cite{Dir} schemes.

\medskip 
As a second-order differential equation in partial derivatives, the Gribov ambiguity equation (\ref {p. con}) would involve two initial conditions in a given time instant $t_0$. 

Just the Gribov ambiguity equation (\ref {p. con}) with mentioned two initial conditions to this equation are responsible indeed for manifest superfluid properties of the Minkowskian Higgs model with BPS monopoles quantized in the "fundamental" \cite{Dir} wise, and our task in the present study therefore is to ascertain these initial conditions.

\medskip
This will be the topic of {\it Section 2}.

It will be convenient to subdivide this Section into two subsections.

In {\it Subsection 2.1} we recall some results regarding the Gauss-shell reduction of the Minkowskian Higgs model with vacuum BPS monopole solutions in terms of topological Dirac variables $\hat A^D$ that have been got in the papers \cite{LP2,LP1,Pervush2}.

Additionally, arguments in favour that these Dirac variables are gauge invariant will be adduced.

\medskip
In {\it Subsection 1.2} we proceed immediately to the analysis of the Gribov ambiguity equation (\ref {p. con}) (responsible for manifest superfluid properties of the Minkowskian BPS monopole vacuum guantized by Dirac) and the initial conditions to this equation.

We demonstrate (this fact was noted as early as in Ref. \cite{LP2}) that the one of these initial conditions is just the fixed Coulomb covariant gauge for Dirac variables $\hat A^D$ in the zero topological sector of the Minkowskian Higgs model with vacuum BPS monopoles (with its "Gribov copies" in other topological sectors of that model).

The second of these initial conditions comes to resolving the Gribov ambiguity equation (\ref {p. con}) in terms of vacuum BPS monopole solutions in the YM sector and perturbation excitations over this BPS monopole vacuum. 

As it was noted in \cite{LP2,LP1}, mentioned excitations $\bar A_i({\bf x},t)$ belong to the class of {\it multipoles} and have the behaviour $1/r^{l+1}$ ($l>0$) at the spatial infinity. 

Thus the Cauchy problem to the Gribov ambiguity equation (\ref {p. con}) in the fixed time instant $t_0$ will be formulated.

\medskip
The important property of the Gribov ambiguity equation (\ref {p. con}) we shall encounter in Subsection 2.2 is that it characterizes the Minkowskian BPS monopole vacuum suffered the Dirac fundamental quantization \cite{Dir} as an incompressible liquid possessing additionally the superfluiduty. 

In demonstrating this property of the Minkowskian BPS monopole vacuum we follow the arguments \cite{gidrodinamika}.

As it is well known from hydrodynamics \cite{gidrodinamika}, the continuity equation 
$$ \frac{\partial \rho }{\partial t}+ {\rm div} \rho {\bf v}=0$$
(with $\rho$ and ${\bf v}$ being, respectively, the density and velocity of the considered liquid) is simplified in a radical way as $\rho = {\rm const}$ (i.e. when the density remains constant along the whole volume besetting by the liquid during the whole time of motion). \par
The  condition $\rho = {\rm const}$ means just that the liquid is {\it incompressible} \cite{gidrodinamika}.
 
In  this case the continuity equation acquires the  simplest look 
$$  {\rm div}~ {\bf v}=0
       $$
or
$$   \Delta \phi=0      $$
if ${\bf v}={\rm grad}~\phi$ for a scalar field $\phi$, i.e. if the considered liquid is potential.

Thereafter, it should be recalled \cite{rem1} that the vacuum "magnetic" field $\bf B$ plays the same role that the (critical) superfluiduty velocity ${\bf v}_0$ of the superfluid component in a liquid helium II \cite{N.N.}.

Then one comes to the Gribov ambiguity equation (\ref {p. con}) instead Eq. $\Delta \phi=0 $ upon replacing
$$ {\bf v}_0\Leftrightarrow {\bf B}; \quad   \phi \Leftrightarrow  \Phi.              $$
Indeed, the property of the Minkowskian BPS monopole vacuum quantized by Dirac to be an incompressible liquid is reduced to the existence of definite topological invariants  characterized this vacuum.

There are the magnetic charge ${\bf m}$ and the degree of the map referring to the \linebreak $U(1)\subset SU(2)$ embedding.

These topological invariants  will be us discussed in detail in Subsection 1.2.

\bigskip
In {\it Section 2} we study the properties of stationary Gribov topological multipliers $ v^{(n)}({\bf x})$ entering topological Dirac variables $A^D$ in the fixed time instant $t_0$.

Indeed, as it was demonstrated in Ref. \cite{David2},
$$ v^{(n)}({\bf x})= U(t,{\bf x}) \vert_{t=t_0};      $$
herewith the matrices $ U(t,{\bf x}) $ are functions of the BPS monopole background \cite{LP1}.

As to Gribov topological multipliers $ v^{(n)}({\bf x})$, their outlined look was elucidated in Refs. \cite{LP2, LP1}:
$$v^{(n)}({\bf x})= \exp (n \hat \Phi({\bf x})),$$ 
with $\hat \Phi({\bf x})$ being the {\it Gribov phase}, a scalar value may be expressed through a combination of the Pauli matrices $ \tau^a$ ($a=1,2,3$) and a Higgs BPS monopole.

\medskip
The principal result will be got in Section 2 is to show the spatial asymptotic
 \be \label{graniza}  v^{(n)}({\bf x})\to \pm 1\quad {\rm as}~ \vert {\bf x}  \vert \to \infty    \ee
for Gribov topological multipliers $ v^{(n)}({\bf x})$. 

We shall follow the paper \cite {Azimov} at grounding this fact.

More precisely,  it will be shown that Gribov topological multipliers $ v^{(n)}({\bf x})$ may be rewritten to depend on Euler angles $\phi_i$ ($i=1,2,3$).   These, in turn, would be chosen in such a wise that the boundary condition (\ref{graniza}) is satisfied.

As it was demonstrated in \cite {Azimov} (see also Ref. \cite {fund}),  the spatial asymptotic (\ref{graniza}) ensures the infrared ({\it topological}) confinement of Gribov multipliers $ v^{(n)}({\bf x})$ in fermionic and gluonic Green functions in all the orders of the perturbation theory (the author intend to return to this question in one of his future articles).
\section{Cauchy problem to Gribov ambiguity equation and superfluidity of BPS monopole vacuum.}
\subsection{Constraint-shell reduction of Minkowskian Higgs model in terms of topological Dirac variables.}

In this subsection, playing a rather auxiliary role in the present study, which principal goal is formulating the Cauchy problem to the Gribov ambiguity equation and revealing its importance for the superfluid effects occurring inside the BPS monopole vacuum, we recall the said in the papers \cite{LP2,LP1,Pervush2} about the constraint-shell reduction of the Minkowskian Higgs model with vacuum BPS monopole solutions in terms of topological Dirac variables \cite{David2, David3}.

\medskip
The base of the Dirac fundamental approach \cite{Dir} to the quantization of the Minkowskian Higgs model (with vacuum BPS monopoles) is solving the YM Gauss law constraint
\be
\label{Gauss}
\frac {\delta W}{\delta A^a_0}=0 \Longleftrightarrow [D^2(A)]^{ac}A_{0c}= D^{ac}_i(A)\partial_0 A_{c}^i
\ee
with the covariant (Coulomb) gauge
\be
\label{Aparallel}
A^{a\parallel}\sim [D^{ac}_i(\Phi ^{(0)})A_c^{i ~(0)}]=0\vert _{t=0}
\ee
(involving the YM BPS monopole background $\Phi ^{(0)}$ in the zero topological sector of the Minkowskian Higgs model with vacuum BPS monopoles).

Eq. (\ref {Aparallel}) permits the transparent treatment as the absence, in the initial  time instant $t_0$, of longitudinal components of YM fields.

Meanwhile, temporal components $A_0^a$ of YM fields, standing on the left-hand side of the Gauss law constraint (\ref{Gauss}), are, indeed, nondynamical degrees of freedom, the quantization of which contradicts to the
quantum principles \footnote{More precisely, the non-dynamic status of $A_0^a$ is not compatible with the
 quantization of these fields via their
 definite fixing  (e.g. via the YM Gauss law constraint (\ref{Gauss})), while  the appropriate zero canonical momenta
$$E_0\equiv\partial {\cal L}/\partial (\partial_0
 A_0)=0$$
contradict  the commutation relations and uncertainty principle  \cite{Pervush2}.}, can and would be removed following \cite{Dir}.

At attempting \cite{Pervush2} to solve the YM Gauss law constraint (\ref{Gauss}) in terms of nonzero stationary initial data:
\be
\label{init}
\partial _0 A_i ^c =0 \Longrightarrow  A_i ^c(t,{\bf x})=\Phi_i^{c(0)}({\bf x}):
\ee
the YM vacuum BPS monopole solutions in the Minkowskian Higgs model (in the zero topological sector of that model) is one of examples
  resolving (\ref{init}) the YM Gauss law constraint (\ref{Gauss}), the former acquires the look
\be
\label{Gauss2}
\partial _0 [D^{ac}_i(\Phi ^{(0)})A_c^{i~ (0)}]=0
\ee 
upon the removal, ala   \cite {Dir}, temporal YM components $A_0^a$ from the left-hand side of this Eq.

In this case the covariant Coulomb gauge (\ref{Aparallel}) just will be \cite {Pervush2} the solution to the YM Gauss law constraint (\ref{Gauss2}) in the fixed time instant $t_0$ at resolving (\ref{init}) this constraint in terms of YM vacuum BPS monopole solutions.

In particular, vacuum YM BPS monopole solutions (\ref{init}) to the YM Gauss law constraint (\ref{Gauss}) may be chosen to be transverse and satisfy (\ref{Aparallel}).

Note, in connection with the said, that such possibility for vacuum YM BPS monopole solutions to be transverse fields is the specific of rather the Dirac fundamental quantization method \cite {Dir} (involving resolving the YM Gauss law constraint (\ref{Gauss})) than the "heuristic" one \cite {FP1}.

As it was discussed in \cite {rem1,fund} (repeating the arguments \cite {Al.S.}), in Minkowskian Higgs models with monopoles quantized in the "heuristic"  \cite {FP1} wise, it is enough only to fix the temporal (Weyl) gauge $A_0=0$ via the $\delta(A_0)$ multiplier in appropriate FP path integrals. 
\medskip
With taking account of the covariant Coulomb gauge (\ref{Aparallel}), the YM Gauss law constraint
 (\ref{Gauss2}) acquires the alternative look \cite {LP1}
\be \label{Aparallel1} \partial _t A^{a\parallel}[A_c^{i(0)}(t,{\bf x})] =0\ee
over the set of (topologically trivial) vacuum YM BPS monopole solutions (\ref{init}).

On the other hand, Eq. (\ref{Aparallel1}) may by treated also as the covariant Coulomb gauge. Just such treatment of (\ref{Aparallel1}) we shall utilize in the present study.

\medskip
It is easy to see that Eqs. (\ref{Gauss2}) and (\ref{init}) are mathematically equivalent (more exactly, assuming (\ref{init}), one comes to (\ref{Gauss2}); as well as in another cases resolving the YM Gauss law constraint (\ref{Gauss}) though, since the covariant derivative $D$ and the time one, $\partial_0$, are commutative.). This remark will play the crucial role in the near future.

\medskip
Indeed, the YM Gauss law constraint (\ref{Gauss}) refers to the "pure" YM theory, without another (quantum) fields: for instance, Higgs and fermionic modes.

Meanwhile, if enumerated fields are present in a gauge non-Abelian model, it may be demonstrated (see e.g. \S 15 in \cite {Gitman}) that the Gauss law constraint would contain  a current item $\rho^L$ that is, in the non-Abelian theory,  the sum of   two items: the non-Abelian and fermionic  currents. 

Regarding Higgs fields, also contributing to the current item $\rho^L$ \cite{Gitman}, we should like to note the following.

When, at going over to the Minkowski space and violating the initial $SU(2)$ gauge symmetry group down to its (stationary) $U(1)$ subgroup, Higgs modes appear, having herewith look of stationary vacuum solutions, monopoles, and, perhaps, perturbation excitations over this monopole vacuum of higher orders (such that one can neglect them in the "classical" YM Hamiltonian formalism), there are no an essential contribution from Higgs modes to the YM Gauss law constraint, at least in the lowest order of the perturbation theory.

\medskip
The particular case resolving the YM Gauss law constraint (\ref{Gauss}) with the covariant Coulomb gauge is its resolving in terms of topological Dirac variables \cite {David2, David3, Pervush2}: transverse and gauge invariant functionals of YM fields.

With account of the Gribov topological degeneration \cite{Gribov} of non-Abelian data, topological Dirac variables \cite {David2, David3, Pervush2}, satisfying the covariant Coulomb gauge (\ref{Aparallel}) in the zero topological sector of the Minkowskian Higgs model quantized by Dirac \cite {Dir} (and involving vacuum BPS monopole modes) and its "Gribov copies" \cite 
{Pervush2}
\be
\label{transv}
D_i^{ab} (\Phi _k^{(n)}){A}^{i(n)}_b =0, 
\ee 
in topologically nontrivial ($n\neq 0$) sectors of that model,  have the shape \cite {LP1}
\bea
\label{degeneration}
\hat A_k^D = v^{(n)}({\bf x})T \exp \left\{\int  \limits_{t_0}^t d {\bar t}\hat A _0(\bar t,  {\bf x})\right\}\left({\hat A}_k^{(0)}+\partial_k\right ) \left[v^{(n)}({\bf x}) T \exp \left\{\int  \limits_{t_0}^t d {\bar t} \hat A _0(\bar t,{\bf x})\right\}\right]^{-1}, \eea 
with  the symbol $T$  standing for   time ordering  the matrices under the exponent sign. \par
In the initial time instant $t_0$, the topological degeneration of initial (YM) data
comes thus to  "large" stationary matrices $v^{(n)}({\bf x})$ ($n\neq 0$) \footnote{The terminology "large" originates from the papers \cite {Jack}.  Following  \cite {Jack}, we also shall refer to the topologically trivial matrices $v^{(0)}({\bf x})$ as to the "small" ones.} depending on  topological numbers $n\neq 0$ and called the factors of the
Gribov topological degeneration or simply the \it Gribov multipliers\rm.  \par

One attempts \cite {LP2,LP1, Pervush2}  to find Gribov multipliers $v^{(n)}({\bf x})$, belonging to the $U(1)\subset SU(2)$ embedding in the Minkowskian Higgs model, as
$$ \exp [n \hat\Phi _0({\bf x})],$$
implicating the {\it Gribov phase} $\hat\Phi _0({\bf x})$.\par
It will be demonstrated in Section 2 (repeating the arguments \cite {David3})  that $\hat\Phi _0({\bf x})$ is a scalar constructed by contracting the Pauli matrices $\tau^a$ and Higgs vacuum BPS monopole modes. 

More exactly, in the initial time instant $t_0$, the topological Dirac variables (\ref{degeneration}) acquire the look
\be
\label{degeneration1}
{\hat A}^{(n)}_k= v^{(n)}({\bf x}) ({\hat A}_k^{ (0)}+
\partial _k)v^{(n)}({\bf x})^{-1},\quad v^{(n)}({\bf x})=
\exp [n\Phi _0({\bf x})].
\ee 

\medskip
Regarding exponential multipliers (with the braces) in (\ref{degeneration}), it may be noted the following. \par
In the Minkowskian Higgs model with  vacuum BPS monopole solutions, it is quite logical to assume that these would depend explicitly 
on  YM BPS monopole modes (belonging to the zero topological sector of that Minkowskian Higgs model) \cite {LP1}. \par
Thus one can always consider the matrices
\be \label{dress}
U(t,{\bf x})= v({\bf x})T \exp \{\int  \limits_{t_0}^t 
[\frac {1}{D^2(\Phi^{\rm BPS })} \partial_0 D_k (\Phi^{\rm BPS }){\hat A}^k]~d\bar t ~\}. \ee
Following the work  \cite{LP1}, let us denote as $U^D[A]$ the exponential expression in (\ref{dress}); this expression may be rewritten \cite{LP1,Pervush1} as
\be
\label{UD}
U^D[A]= \exp \{\frac {1}{D^2(\Phi^{BPS })} D_k (\Phi^{BPS }) {\hat A}^k\}
\ee
over the stationary BPS monopole background.

We shall refer to the matrices $U^D[A]$ as  to the \it Dirac \rm "\it dressing\rm" of non-Abelian \linebreak fields (following \cite{LP1,Pervush2}). 

\medskip
Meanwhile, temporal component of topological Dirac variables $\hat A^D$ would be removed (following Dirac \cite{Dir}) according to the grounds us stated above: these grounds come to the nondynamical status of temporal components of YM fields.

Thus \cite{David2}
\be
\label{udalenie}
U(t,{\bf x}) (A_0^{(0)}+\partial_0) U^{-1}(t,{\bf x})=0.
\ee
Eq. (\ref {udalenie}) can serve for specifying {\it Dirac } matrices $ U(t,{\bf x})$.

\medskip
In order for Dirac variables (\ref{degeneration}) to be gauge invariant, it is necessary \cite{David2,Azimov} that exponential multipliers $ U(t,{\bf x})$, (\ref{dress}), entering (\ref{degeneration}), cancel the action of YM gauge transformations 
\be
\label{gauge}
{\hat A}^u_i=u(t; {\bf x})({\hat A}_i+\partial_i)u^{-1}(t,{\bf x}).\ee
The said may be written down as the transformations law for $ U(t,{\bf x})$:
\be
\label{z-n dlja U}
U(t,{\bf x})\to U_u (t,{\bf x}) =  u^{-1}(t,{\bf x}) U(t,{\bf x}).
\ee 
If the transformations law (\ref{z-n dlja U})
for matrices $ U(t,{\bf x})$ takes place, it is easy to demonstrate that (topological) Dirac variables (\ref{degeneration}) are indeed gauge invariant performing the following computations proposed in Ref. \cite{Azimov}.

Denoting matrices $ U(t,{\bf x})$ as $v[A]$, one write \cite{Azimov}
\be \label{proverca}
 \hat A_i^D[A^{u}]= v[A]~ u^{-1} u(\hat A_i+   \partial_i) u^{-1} u ~v[A] ^{-1} = \hat A_i^D \ee
(the fact that matrices $v[A]$ and $u$ are commute with each other was utilized at the computations (\ref{proverca}); for the $U(1)\subset SU(2)$ embedding taking place in the Minkowskian Higgs theory in question, this is obviously and is not associated with additional difficulties).

\bigskip In the light of Eqs. (\ref{degeneration}) and  (\ref{proverca}), the following can be concluded, and this is very important and interesting. Eq. (\ref{degeneration}) 
seems to be rather a kind of a topological map from the zero topological sector of the model we study now to its $n$$^{\rm th}$ topological sector. Herewith its zero topological sector is represented by gauge fields $\hat A_k^{(0)}$.

There exists a simple mathematic model (see Lecture 2 in \cite{Postn4}) which  describes correctly such a topological map. This is the {\it covering construction} where the set $B$ of gauge (covariant) fields  $\hat A_k^{(0)}$ constitutes the base of such a 
 covering while its discrete infinitely-valent fibre is the set of all fields $\hat A_k^{D(n)}$ ($n \in {\bf Z}$), the (topological) Dirac variables (\ref{degeneration}).

\bigskip
Finishing this subsection, we should like make some concluding remarks regarding properties of topological Dirac variables.

\medskip
1. In the majority of formulas we have encountered in the present subsection, YM fields $\hat A$ are present.

These  fields have indeed the following look in the Planckian $\hbar$, $c$ units \cite{Pervush3}:
\be \label{hatka}
\hat A_\mu = g \frac {A_\mu^a\tau _a}{2i \hbar c}.  
\ee
The account of $\hbar$ and $c$ in latter Eq. is closely connected with the actual value $g/(\hbar c)$ of the strong coupling constant.

\medskip
2. The covariant Coulomb gauge (\ref{Aparallel1}) for YM fields acquires its look \cite{David2, Azimov, Pervush3} 
\be \label{Aparallelr}
D_i^{ab} (\hat A^D) \partial_0 (\hat A_i^D)\equiv 0
\ee 
in terms of topological Dirac variables $ \hat A^D$.

Moreover, the Coulomb gauge (\ref{Aparallel}), (\ref {transv}) for topological Dirac variables $\hat A^D$ \cite{rem1,fund}:
$$ D_i \hat A^{Di}=0, $$
and the removal (\ref{udalenie}) \cite{David2} of temporal components of these topological Dirac variables imply the generalized Lorentz gauge
\be \label{Lorentz}
D_\mu \hat A^{D\mu}=0
\ee
for them.

\medskip
3. It is quite naturally to expand any (say, transverse) YM field $\hat A({\bf x},t)$, (\ref{hatka}), in the sum of a background {\it stationary} (vacuum) field $\hat \Phi({\bf x})$ and a field $\hat {\bar A}({\bf x},t)$ belonging to the excitation spectrum over the background $\hat \Phi$:
\be \label{suma} 
\hat A({\bf x},t) =\hat \Phi({\bf x}) +\hat{\bar A}({\bf x},t).
\ee
In the said is the essence of the postulate has been suggested in Ref. \cite{Pervush1} for Minkowskian non-Abelian theories.

In  Minkowskian non-Abelian theories, physical (transverse) YM variables always may be represented as   sums
of the (singular) stationary Bose condensate $\underline b({\bf x})$ and dynamical regular fields \rm $\underline a({\bf x},t)$, treated as perturbation excitations over this  (singular) stationary  Bose condensate: 
\be 
\label{postulat2} A({\bf x},t)= \underline b({\bf x})+ \underline a({\bf x},t).
\ee
Generally speaking, this assumption is irrelevant to choosing the space where a non-Abelian gauge theory is considered (either the Minkowskian or Euclidian one), but in Minkowskian non-Abelian models it has by far interesting consequences than in Euclidian non-Abelian models.

Extracting the c-number field $\underline b({\bf x})$, one should ensure  herewith  that  energies of (stationary) quantum states are finite  and that these quantum states are stable. \par
 The decomposition  (\ref{postulat2}), in the Minkowski space, 
shouldn't be suppressed by factors of the $ \exp (-S_E(\underline b)/\hbar)$  type, always taking place in the  Euclidian space $E_4$. And  it is the one  more argument in favour of going over to the Minkowski space from the Euclidian one at  considering   the non-Abelian vacuum (in particular, that quantized by Dirac \linebreak \cite{Dir}) \footnote{Principal shortcomings of the  Euclidian instanton non-Abelian model \cite{Bel} were pointed out recently in the paper \cite{fund} (repeating the arguments of Refs. \cite{Pervush2, Pervush1, Arsen}).\par 
Actually, as a principal shortcoming of the Euclidian instanton non-Abelian model \cite{Bel}, the purely imaginary values of the topological momentum \cite{Pervush2, Pervush1, Arsen}
$$ P_{\cal N}=\pm 8\pi i/g^2\equiv 2\pi k + \theta,         $$
referring to the Euclidian $\theta$-vacuum
(with $\theta\in [-\pi,\pi]$ \cite{Pervush1}), may be indicated.

This involves \cite{fund} the bad behaviour of the $\theta$-vacuum
plane wave function \rm \cite {Pervush1}
$$  \Psi _0[A]=\exp (iP_{\cal N}X[A])         $$
(implicating the winding number functional $ X[A]$ taking integers) at the minus sign before $P_{\cal N}$.

As a result, it is impossible to give the correct probability description of the instanton $\theta$-vacuum; that is why the latter one refers  to unobservable, i.e. {\it unphysical}, values.

In the paper \cite{Arsen}, the effect appearing the purely imaginary values of the topological momentum $ P_{\cal N}$ for the Euclidian $\theta$-vacuum was referred to as the so-called \it no-go theorem\rm: the absence of physical solutions in the Euclidian 
instanton YM (non-Abelian) theory \cite{Bel}.}.

\medskip
In the Minkowskian Higgs model quantized by Dirac, the postulate (\ref{postulat2}) \cite {Pervush1} takes the shape \cite {Pervush2} 
\be
 \label{s12}
 \hat A_i^{ (n)}(t,{\bf x}) = \hat\Phi_i^{ (n)}({\bf x}) + {\hat{\bar A}}_i^{ (n)} (t,{\bf x})
\ee
in each topological class of that model.

If this model includes vacuum BPS monopole modes, $\hat\Phi_i^{ (n)}({\bf x}) $ are just BPS monopole modes  belonging to the YM sector and satisfying Eq. (\ref{init}).

 Eq. (\ref{s12}) is in a good agreement with Eqs. (\ref{degeneration}), (\ref{degeneration1}) for topological Dirac variables $\hat A^D$. It is, actually, the look of $\hat A^D$ meeting the postulate  (\ref {postulat2}) \cite{Pervush1}. 

In particular, topological Dirac variables $\hat A_i^{ (n)}(t,{\bf x})$, (\ref {s12}), are transverse and gauge invariant functionals of YM fields, satisfying herewith the covariant Coulomb gauge (\ref {transv}) in each topological class of the Minkowskian Higgs model quantized by Dirac \cite{Dir}.

Furthermore, there may be assumed (as it was done in \cite{Pervush2}) that fields ${\hat{\bar A}}_i^{ (n)} (t,{\bf x})
$ belong to the class of {\it multipoles}, with their $O(1/r^{l+1})$ ($l>0$) behaviour at the spatial infinity.

Comparing Eqs. (\ref {s12}) and (\ref {degeneration1}), one can single out \cite{Pervush2} the topologically degenerated BPS monopole background 
\be
 \label{mon.deg}
 {\hat \Phi_i} ^{(n)}:= v^{(n)}({\bf x})[{\hat \Phi_i} ^{(0)}+\partial _i]v^{(n)}({\bf x})^{-1},\quad v^{(n)}({\bf x})=
\exp [n\hat \Phi _0({\bf x})],
 \ee
 and  topologically degenerated multipoles:
 \be
 \label{mult}
 {\hat {\bar A}}^{(n)}:= v^{(n)}({\bf x}){\hat {\bar A}}^{(0)}v^{(n)}({\bf x})^{-1},
\ee
as excitations over this BPS monopole background. 

Emphasise again that the fields ${\hat \Phi_i} ^{(n)}$ and ${\hat {\bar A}}^{(n)}$ are transverse functionals of YM fields.

Topological Dirac variables (\ref{mon.deg}), (\ref{mult}) permit the transparent physical interpretation as YM modes ``dressed'' in the Higgs Bose condensate given in the model \cite{LP2,LP1,David2,David3,Pervush2} in the shape of Higgs BPS monopoles.
\subsection{Cauchy problem to Gribov ambiguity equation and superfluid properties of Minkowskian Higgs model with BPS monopoles quantized by Dirac. }
As it was demonstrated in Ref. \cite{LP2,LP1,Pervush2}, the Coulomb constraint-shell gauge (\ref{transv}) keeps its look in each topological class of the  Minkowskian Higgs model with vacuum BPS monopole solutions quantized by Dirac \cite{Dir} if the  Gribov phase $\hat\Phi _0({\bf x})$, entering Eqs. (\ref {degeneration}), (\ref {degeneration1}) for topological Dirac variables in that model, satisfies \it the equation of the Gribov ambiguity \rm  (or simply the \it Gribov equation\rm)  
\be
\label{Gribov.eq} [D^2 _i(\Phi _k^{(0)})]^{ab}\Phi_{(0)b} =0.
\ee 
The origin of latter Eq. is in the standard definition of a "magnetic"  field, 
\be \label {magnet} B_i^a= \epsilon_{ijk} (\partial^j A^{ak} +\frac {g}{2}\epsilon ^{abc}A_{b}^j A_{c}^k).\ee
Really, the values $D_i A^{ia}$ (in particular, $D_i A^{i D}$ if topological Dirac variables $A^D$ are in question) have the same dimension  that a "magnetic" YM field $ B_i^a$, given via (\ref{magnet}).

Then it is easy to see that the Gribov ambiguity equation (\ref{Gribov.eq}) is the consequence of the Bogomol'nyi equation (\ref{Bog}), implicating (topologically trivial) Higgs vacuum BPS monopole modes $\Phi_{(0)}$. 

 Speaking about the connection between the Bogomol'nyi and Gribov ambiguity equations, note that this connection may be given via the Bianchi identity
$$\epsilon ^{ijk}\nabla _i F_{jk}^b =0,$$ 
that is equivalent to 
$$ D B=0$$
in terms of the (vacuum) "magnetic"  field $\bf B$, (\ref{magnet}).

\medskip
Meanwhile, mathematically, the Gribov equation (\ref{Gribov.eq}) implies that the FP determinant \cite{Gitman,Baal}
\be 
\label{FP}
{\rm det}~ (\hat \Delta ^b_a) \equiv {\rm det}~ (-D ^b_{ai}\partial^i) = {\rm det}~ (-[\partial_i^2 + \partial_i~ {\rm ad} (A^i)]),
 \ee 
with
$$ {\rm ad} (A) X \equiv [A,X]$$ 
for an element $X$ of the $SU(2)$ Lee algebra, becomes  zero at setting the constraint-shell (Coulomb) transverse gauge (\ref{Aparallel}), (\ref {transv}) in the Minkowskian Higgs model quantized by Dirac \cite{Dir}.

\medskip
In the terminology of the paper \cite{Baal} (see also \cite{fund}), there was shown that the $L^2$ (Lebesgue)  norm  $\Vert A \Vert ^2$ of an YM potential $A$ along the fixed gauge orbit, i.e. 
\be 
\label{Lebeg} 
\Vert A \Vert ^2 \equiv F_A (g)  = -\int \limits_ M d^3x ~{\rm tr} [( v^{-1}(x) A_i v(x) + v(x) \partial_i v^{-1}(x))^2], 
\ee 
with $ v(x) \in SU(2)$,
attains its (local) minimum as this YM potential is transverse,  \linebreak $ \partial_i A^i=0$; in 
this case $\hat \Delta$, (\ref{FP}), 
 becomes a positive defined operator \footnote{ In the paper \cite {Baal}, regarding the Euclidian instanton non-Abelian theory \cite {Bel}, the { Weyl} gauge $A_0=0$, has been fixed in the theory \cite {Bel} and just resulting instanton stationary solutions, was utilized. This implied expressing the operator $\hat \Delta$ in terms of only spatial indices.

The same result is achieved at the Gauss-shell reduction of the Minkowskian Higgs model (with vacuum BPS monopole solutions), us discussed in the previous subsection, with ruling out (\ref {udalenie}) \cite {David2} temporal  components of topological Dirac variables $A^D$. }.

The set of all such  YM potentials is called \it  the Gribov region\rm. We shall denote it as  $\Omega$ following \cite{Baal}, while its boundary $\partial \Omega$ is called \it  the Gribov horizon \rm \cite{Baal}.  

At the Gribov horizon $\partial \Omega$, the lowest eigenvalue of the FP operator $\hat \Delta$, (\ref {FP}), vanishes, and points on $\partial \Omega$ are associated with coordinate singularities \footnote{In some physical literature, e.g. \cite{Gribov, Sobreiro}, the equation
$$ (\hat \Delta (\Psi)\equiv  (-\partial_i D^i (A) (\Psi))) = -( \partial_i^2  \Psi + \partial_i~ {\rm ad} (A^i)) \Psi = \epsilon ~(A)\Psi )$$ 
is treated as a specific Schr\"{o}dinger equation, with $A_{i}
$ playing the role
of a potential.

For small values of $A_{i}$ (that now don't assumed to be transverse, $\partial_i A^i\neq 0$), this equation is solvable for positive $\epsilon $ only. \par
More
precisely, denoting by $\epsilon_1(A), \epsilon_2(A),
\epsilon_3(A), \dots $ the eigenvalues corresponding to a given
field configuration $A$, one has that, for small $A_{k }$,
all the $\epsilon_i(A)$ are positive, $\epsilon_i(A)>0$. \par
However, for a sufficiently large value of the field $A_{\mu}$,
one of the eigenvalues, say $\epsilon_1(A)$, turns out to vanish,
becoming then negative as the field increases further, and so on. \par
As in the case
of the Schr\"{o}dinger equation, this means that the field of the Schr\"{o}dinger equation, this means that the field
$A_{\mu}$ is large enough to ensure the existence of negative
energy solutions, { i.e.} bound states. \par
For a greater magnitude
of the field $A_{\mu}$, a second eigenvalue: say, $\epsilon_2(A)$, will vanish, becoming then negative as the field increases again. \par
Following Gribov \cite{Gribov}, one can thus subdivide the functional space of gauge fields into the regions $C_{0}, C_{1}, C_{2}, \dots, C_{n}$ over which the  FP operator $\hat \Delta $,  (\ref{FP}), has, respectively,  $0, 1, 2, \dots, n$ negative eigenvalues. \par
These regions are separated by lines $l_{1}, l_{2},
l_{3},\dots , l_{n}$ on which the  FP operator $\hat \Delta $ takes its zeros. \par
More exactly, in the region $ C_{0}$  all the eigenvalues of the FP operator $\hat \Delta $ are positive, i.e. $\hat \Delta >0$.\par
At the boundary  $l_{1}$ of the region $C_{0}$, the first vanishing eigenvalue of the FP operator $\hat \Delta $ appears; namely on $l_{1}$ the FP operator $\hat \Delta $
possesses a normalizable zero mode $\chi$: 
$$ \hat \Delta~\chi=0.$$ 
In the region $C_{1}$, the FP operator $\hat \Delta $ has one bound
state, { i.e.} one negative energy solution. At the boundary
$l_{2}$, a zero eigenvalue reappears. 

In the region $C_{2}$, the FP operator $\hat \Delta $ has two bound states, { i.e.} two
negative energy solutions. On $l_{3}$ a zero eigenvalue shows up again, and so on. \par
Just the boundaries $l_{1}, l_{2},
l_{3},\dots , l_{n}$,  on which the FP operator $\hat \Delta $,  (\ref{FP}), has zero
eigenvalues (the number of which coincides with the number of the boundary) are called the Gribov horizons in the terminology \cite{Gribov}. \par
For instance, the
boundary $l_{1}$, where the first vanishing eigenvalue appears, is
called {\it the first horizon} \cite{Sobreiro}.
The connection between this definition  \cite{Gribov,Sobreiro} of the Gribov horizons and that given in Ref.  \cite{Baal} can be given with the relation
$$ \partial \Omega= \bigcup\limits _i ~ l_i.  $$
\par
\medskip
Note  also \cite{Sobreiro} that the FP operator $\hat \Delta $,  (\ref{FP}), is Hermitian one:
$ \hat \Delta ^{\dagger}= \hat \Delta $.\par 
Roughly, it is associated with the manifest Hermitian operator $i\partial$ (the momentum operator) entering Eq. (\ref{FP})  together with YM potentials $A$, chosen to be real.  \par 
The complete proof that the FP operator $\hat \Delta $,  (\ref{FP}), is indeed Hermitian one is given in the paper \cite{Sobreiro}, we recommend our readers for studying the question.  }. \par 
The next important definition regarding the FP operator $ \hat \Delta $,  (\ref{FP}), and its determinant ${\rm det}~\hat \Delta $ is the \it fundamental  domain \rm \cite{Baal}, we shall denote as $\Lambda$ henceforth. 
It is the set of absolute minima of the norm functional (\ref{Lebeg}). \par 
\medskip
Upon our analysis of the FP operator $ \hat \Delta $,  (\ref{FP}), it becomes obvious that the Gribov horizon $\partial \Omega$ is just the set in the space of topological Dirac variables $A^D_i$ ($i=1,2$), (\ref{degeneration}), transverse functionals of gauge fields $A$, over which the Gribov ambiguity equation (\ref{Gribov.eq}) is satisfied. \par
In the Minkowskian Higgs model quantized by Dirac \cite{Dir} and involving vacuum BPS monopole modes, the Gribov horizon $\partial \Omega$ may be expressed alternatively (in comparison with the above definition given in Ref. \cite{Baal}) as the set of (topologically degenerated) Higgs vacuum BPS monopole modes $\Phi^{(n)}_a $ belonging to the kernel of the FP operator $\hat \Delta$, (\ref{FP}):
  \be 
\label {ker}
   \Phi^{(n)}_a\in {\rm ker}~ \hat \Delta, \quad n \in {\bf Z}.
\ee 
Actually, it is the set of all the Higgs vacuum BPS monopole modes available in the quested Minkowskian Higgs model quantized by Dirac. \par
From the definition \cite{Baal} of the Gribov region $\Omega$ it follows that this region is swept by the family  of  Coulomb covariant gauges (\ref{Aparallel}) for topological Dirac variables $A^D$, in the zero topological sector of the Minkowskian Higgs model quantized by Dirac, and their  Gribov  copies (\ref {transv}) in other  topological sectors of that model. \par
\medskip
Moreover, in terms of the definitions of the Gribov region $\Omega$ and its boundary, the Gribov horizon $\partial \Omega$, given in  the papers \cite{Baal, Sobreiro}, the said implies that the Gribov horizon $\partial \Omega$ becomes the 
space-like surface $\partial_i^2=0$ when the constraint-shell transverse gauge (\ref{transv}) for topological Dirac variables $A^D$ is fixed course the Gauss-shell reduction  of the Minkowskian Higgs model with vacuum BPS monopoles.
\par
Further, the Dirac removal (\ref{udalenie}) \cite{David2} of temporal YM components implies formally that $\partial_ \mu \hat A_\mu ^D=0$. \par
 Then in terms of topological Dirac variables (\ref{degeneration}), (\ref{udalenie}), satisfying the constraint-shell transverse gauge (\ref{transv}), the Gribov horizon $\partial \Omega$ may be continued actually to the isotropic surface $\partial_\mu^2=0$, and this means that gauge fields in the Minkowskian Higgs model quantized by Dirac \cite{Dir}, taking the look of (topological) Dirac variables (\ref{degeneration}), (\ref{udalenie}), upon the constraint-shell reduction of this theory, are, indeed, massless fields. 
\par  
Eq. $\partial_\mu^2=0$ for the Gribov horizon $\partial \Omega$ in the Minkowskian  Higgs model quantized by Dirac implies  (cf. \cite{Sobreiro}) that there is only a one Gribov region $C_0$ in that theory, where the FP operator (\ref{FP}) possesses 0 negative eigenvalues. 
\par 
Respectively, instead of the set $\{l_i\}$ \cite{Sobreiro} of Gribov horizons in an YM model where the transverse gauge of fields isn't fixed, now we encounter only the one Gribov horizon $\partial \Omega$ \footnote{ Indeed, the "gauge" $\partial_i A^i=0$, utilized in Ref. \cite{Baal} is "narrower" than the transverse gauge $D_i A^i=0$ (the Coulomb covariant gauge (\ref{Aparallel}), (\ref {transv}) for topological Dirac variables $A^D$, (\ref{degeneration}), in the Minkowskian Higgs model quantized by Dirac is the particular case of such transverse gauge). \par  
But one can think that the transverse gauge $D_i A^i=0$ comes to the gauge $\partial_i A^{ia}=0$ and to the condition
$$  \epsilon_{ijk} A^{ja}A^k_a=0    $$ 
(where the group indices $a$ are written down explicitly). \par  
Just at latter two assuming,  topological Dirac variables $A^D$, (\ref{degeneration}), in the Minkowskian Higgs model quantized by Dirac \cite{Dir} may be treated as massless gauge fields, as it will be discussed below.}.
\par  
On the other hand, such Gribov horizon $\partial \Omega$ possesses, indeed, an infinite set of zero eigenvalues for the FP determinant $\rm det$ $\hat \Delta$ due to the nontrivial cohomological structure of YM fields, topological Dirac variables $\hat A ^D$, has been revealed in the recent papers \cite{LP2}.\par  
The said is also the specific of the Dirac fundamental quantization \cite{Dir} of the \linebreak Minkowskian  Higgs model, involving its Gauss-shell reduction  in terms of Dirac variables (\ref{degeneration}), (\ref{udalenie}).\par  
\medskip
On the face of it, the zero FP determinant,  (\ref{FP}), can involve  nontrivial Dirac 
dressing matrices (\ref{UD}) at automatic fixing the Coulomb gauge  (\ref{Aparallel}) for  Dirac variables (\ref{degeneration}).\par  
But in the initial time instant $t=t_0$ the integral in (\ref{dress}) becomes zero; thus  in this  instant the Gribov ambiguity equation   (\ref{Gribov.eq}) does not affect the gauge
transformations  (\ref{degeneration1}), i.e. the nature of Dirac variables, including the Gribov topological degeneration of initial YM data. \par  
In other words,  in each time instant \rm $t$ one can pick out a space-like surface \rm ${\cal H}(t)$  in the Minkowski space-time over which the topological degeneration of initial YM data occurs in the Minkowskian  Higgs model quantized by Dirac \cite{Dir} (and involving vacuum BPS monopole solutions).

This space-like surface is determined, in effect, by the  Gribov ambiguity equation  (\ref{Gribov.eq}).\par
Note  that  Eqs. (\ref{Gauss2}), (\ref{Aparallel})  can be treated as \it Cauchy  conditions \rm for the Gribov ambiguity equation (\ref{Gribov.eq}) in the (initial) time instant $t_0$.\par   
Therefore to specify the 
space-like surface  ${\cal H}(t_0)$ over which the topological degeneration of initial YM data occurs in the Minkowskian  Higgs model quantized by Dirac, one would solve the  Cauchy  problem (\ref{Gribov.eq}) with the  initial conditions (\ref{Gauss2}), (\ref{Aparallel}),  i.e.  in the class of vacuum YM BPS monopoles (\ref{init}) (and observable YM fields, multipoles  ${\hat {\bar A}}^{(n)}$, (\ref{mult}), as perturbation  excitations over this monopole vacuum with the same topological numbers  that  appropriate monopoles),  satisfying herewith the Coulomb gauge  (\ref{Aparallel}) at resolving the YM Gauss law constraint  (\ref{Gauss}) with removing a la Dirac \cite{Dir} temporal components of YM fields. \par 
\medskip
In the light of the said above about the actual property of the Gribov horizon  $\partial \Omega$ in the Minkowskian  Higgs model quantized by Dirac to be an isotropic surface $p_\mu^2=0$ in the Minkowski space-time (in its {\it momentum} representation), it is easy to see that one can choose the space-like surface \rm ${\cal H}(t_0)$  in the Minkowski space-time (in its {\it coordinate} representation) in such a wise that it will cross the  light cone in the coordinate Minkowski space (as, for instance, it is depicted in the monograph \cite{Penrous1}, in Fig. 1.2, i.e. it can be a three-sphere $S^3$ set by Eq. $t_0^2=x^2+y^2+z^2$). \par
The said implies the absence of the nonzero mass scale for gauge (YM) fields in the Minkowskian Higgs model quantized by Dirac \cite{Dir}.

In this is the principal distinction of that from the well-known (Minkowskian) Higgs model \cite{Cheng,Weinbergg}, involving the choice of the Higgs complex $SU(2)$ doublet $\phi=(\phi_1;~ \phi_2)$ in the shape 
\bea
\label{Higgs doublet}
<\phi>_0=\frac{1}{\sqrt 2} \left( \begin{array}{llcl}   0\\ v \end {array}\right),
\eea
where 
$$ <\phi^\dagger~\phi >=v^2/2\quad {\rm with}\quad v=\mu^2/\lambda. $$
On the other hand, the YM model \cite{Cheng,Weinbergg} doesn't assume fixing transverse gauge for YM fields.

Upon defining the new Higgs field \cite{Cheng}
$$ \phi'=\phi-<\phi>_0:$$
(note that this "trick" is in a good agreement with the postulate (\ref{postulat2}) \cite{Pervush1}, so long as by that expansion there is extracted the vacuum $<\phi>_0$ and excitations $\phi'$ over this vacuum in the Higgs field $\phi$) and its substituting in the  Lagrangian density of the Minkowskian  Higgs model, after some maths,  one gets the nonzero mass \cite{Cheng}
\be \label{MA} M_A=\frac{gv}2\ee
for YM fields.
\par 
\bigskip
Since the Bogomol'nyi equation (\ref{Bog}) \cite{rem1, Al.S.,LP2,LP1} describes correctly (see e.g. \cite{rem1}) manifest superfluid properties inherent in the Minkowskian  Higgs model \cite{Al.S.,BPS, Gold} with vacuum BPS monopole solutions, it is easy to guess that the Gribov ambiguity equation  (\ref{Gribov.eq}), got from the Bogomol'nyi equation (\ref{Bog}) with the aid of the Bianchi identity, is responsible for superfluid properties of  the Minkowskian  Higgs model with vacuum BPS monopoles at its Dirac fundamental quantization \cite{Dir}.\par 
Respectively, the superfluidity proper to the Minkowskian  Higgs model with vacuum BPS monopoles quantized by Dirac \cite{Dir} may be described in the framework of  the Cauchy problem to the Gribov ambiguity equation  (\ref{Gribov.eq}).\par 
As we have discussed above, this Cauchy problem to the Gribov ambiguity equation  (\ref{Gribov.eq}) comes  in the fixed (initial) time instant $t_0$ to laying down two initial conditions to this equation. \par 
There are the Coulomb transverse gauge (\ref{Aparallel}) for topological Dirac variables $A^D_i$, (\ref{degeneration}), and finding solutions to the Gribov equation  (\ref{Gribov.eq}) in the shape (\ref {init}) of stationary vacuum BPS monopoles and perturbation excitations over this BPS monopole vacuum: multipoles (\ref {mult}) possessing same topological numbers that appropriate vacuum BPS monopole solutions. \par
The latter condition is mathematically equivalent to Eq. (\ref {Gauss2}).\par
\medskip
In this case choosing a   
space-like surface  ${\cal H}(t_0)$  in the Minkowski space-time over which the Gribov topological degeneration of initial  data occurs in the Minkowskian  Higgs model quantized by Dirac (this surface is specified at solving the Cauchy problem (\ref{Gribov.eq}), (\ref{Aparallel}),(\ref {init})) acquires the following highly transparent interpretation. \par
Any trajectory of the superfluid potential motion inside the Minkowskian non-Abelian vacuum crosses the total set ${\cal H}^T(t)$   of such surfaces  (may be set in the wise us pointed above \cite{Penrous1}) in each fixed  time instant $t$.
\par

\medskip
In spite of the above mathematical compatibility of the Bogomol'nyi equation (\ref{Bog}) and the Gribov ambiguity equation  (\ref{Gribov.eq}), there is the principal distinction between these equations from the point of view choosing the way how to quantized the Minkowskian  Higgs model with vacuum BPS monopole solutions. 

As it was discussed in the recent papers \cite{rem1,fund}, the Bogomol'nyi equation (\ref{Bog}) is quite compatible with the "heuristic" FP \cite{FP1} quantization scheme, coming for various Minkowskian  Higgs models with monopoles to fixing the temporal (Weyl) gauge $A_0=0$ in appropriate FP path integrals.

Herewith the Bogomol'nyi equation (\ref{Bog}) is derived \cite{Al.S.} issuing from the ordinary {\it constrained} Lagrangian of the Minkowskian  Higgs model without performing the Gauss-shell reduction of the appropriate Hamiltonian.

In detail, upon fixing the temporal (Weyl) gauge $A_0=0$ in the Minkowskian  Higgs model \cite{Al.S.,BPS, Gold} with vacuum BPS monopole solutions, the Bogomol'nyi equation (\ref{Bog}) is derived \cite{Al.S.} at evaluating the {\it Bogomol'nyi bound} \cite{rem1,LP2,LP1,fund}
\be 
\label{Emin}
E_{\rm min}= 4\pi {\bf m }\frac {a}{g},~~~~~~~~~~~~\,\, ~~~~~a=\frac{m}{\sqrt{\lambda}}
\ee 
(where $\bf m$ denotes  the magnetic charge)
of the  energy for the given  configuration of YM and Higgs fields.

In this Eq., the "effective" Higgs mass $m/\sqrt{\lambda}$ appears.

It is taken in the so-called BPS (Bogomol'nyi-Prasad-Sommerfeld) limit \cite{Al.S.,BPS,LP2,LP1} 
\be 
\label{lim} 
\lambda\to 0,~~~~~~m\to 0:~~~~~~~~~~ ~~~~~\frac{1}{\epsilon}\equiv\frac{gm}{\sqrt{\lambda}}\not =0 
\ee 
for the Higgs mass $m$ and Higgs selfinteraction constant $\lambda$. 

The parameter $\epsilon$, the typical size of BPS monopoles, arises in latter Eq. 

\medskip
On the other hand, the Bogomol'nyi equation (\ref{Bog}) is compatible also with the Dirac fundamental quantization  \cite{Dir} of the Minkowskian  Higgs model involving vacuum BPS monopole solutions.

This statement may be explained  again by the actual independence the way deriving the Bogomol'nyi equation (\ref{Bog}) on gauge fixing (indeed, upon removing temporal components of YM fields due to imposing the  gauge condition $A_0=0$; in the framework of the FP "heuristic" quantization scheme \cite{FP1} this can be performed by means the multiplier $\delta(A_0)$ in the appropriate path integral \cite{rem1}, while at the Dirac fundamental quantization \cite{Dir} of the Minkowskian  Higgs model with vacuum BPS monopoles temporal components of YM fields can be ruled out by the standard gauge transformation (\ref{udalenie}) \cite{David2}).

This just justifies the logical and mathematical connection between the Bogomol'nyi and Gribov ambiguity Eqs. us discussed above.

\medskip
Unlike the Bogomol'nyi equation (\ref{Bog}), the Gribov ambiguity equation  (\ref{Gribov.eq}) is associated completely with the Dirac fundamental quantization  \cite{Dir} of the Minkowskian  Higgs model with vacuum BPS monopole solutions, involving its Gauss-shell reduction in terms of topological Dirac variables (\ref{degeneration}).

It just specifies the ambiguity in the choice of these variables, gauge invariant and transverse functionals of YM fields.

\medskip 
The Gribov phase $\hat \Phi({\bf x})$ (entering Gribov topological multipliers $ v^{(n)}({\bf x})$), whose explicit look \cite{LP2,LP1,David3} will be us given in the next section, is, indeed, an $ U(1)\subset SU(2)$ isoscalar constructed by contracting (topologically trivial) Higgs vacuum BPS monopole solutions $\Phi _{(0)a}$ and the Pauli matrices $\tau^a$ ($a=1,2,3$).

The said just allows to assert that the Gribov ambiguity equation (\ref{Gribov.eq}) affects the Gribov phase $\hat \Phi({\bf x})$.

\medskip 
In the paper \cite{rem1} there was traced the transparent parallel between the vacuum "magnetic" field $\bf B$, given via the Bogomol'nyi equation (\ref{Bog}), and the critical velocity ${\bf v}_0$ of the superfluid motion in a liquid helium II specimen.

Herewith  there was cited \cite{Landau52} the enough simple relation
\be \label{alternativ} {\bf  v}_0 =\frac{\hbar}{m} \nabla \Phi(t,{\bf r}),
\ee
between this critical velocity ${\bf v}_0$ and the phase $\Phi(t,{\bf r})$ of the helium Bose condensate wave function $\Xi (t,{\bf r})\in C$. 

The latter one may be given as \cite{Landau52} 
\be
\label{Xi1}
 \Xi (t,{\bf r})= \sqrt {n_0(t,{\bf r})}~ e^{i\Phi(t,{\bf r})},
\ee
with $ n_0(t,{\bf r})$ being the number of particles in the ground energy state $\epsilon=0$ and serves as a complex order parameter in the Bogolubov-Landau  model \cite{N.N.} of the liquid helium.\par 

In Eq. (\ref{alternativ}), $m$ is the mass of a helium atom.

\medskip
We see that the Bogomol'nyi equation (\ref{Bog}), specifying the vacuum "magnetic" field $\bf B$ in the Minkowskian  Higgs model with vacuum BPS monopoles, and Eq. (\ref{alternativ}) \cite{Landau52}, specifying the critical velocity ${\bf v}_0$ of the superfluid motion in the Bogolubov-Landau  model \cite{N.N.} of the liquid helium, have the similar look.

This explains   the role of the Bogomol'nyi equation (\ref{Bog}) as the potentiality condition for the Minkowskian  BPS monopole vacuum.

Really, any potentiality
condition may be written down as \cite{rem1} 
\be \label{potcon} {\rm rot}  ~{\rm grad} ~{ \Phi}=0 \ee
for a scalar field $\Phi$.\par
Thus any  potential field may be represented as ${\rm grad} ~{ \Phi}$ (to within a constant): in particular, this is correctly for the vacuum "magnetic" field $\bf B$ in the Minkowskian  Higgs model with vacuum BPS monopoles. \par
\medskip 
The Gribov ambiguity equation (\ref{Gribov.eq}), following from the Bogomol'nyi equation (\ref{Bog}) due to the Bianchi identity, also may be treated as the potentiality
condition for the Minkowskian  BPS monopole vacuum, but now at the Dirac fundamental quantization \cite{Dir} of the Minkowskian  Higgs model.
\medskip

The next consequence of  the Bianchi identity $D~B=0$ is that the Minkowskian  BPS monopole vacuum may be considered as an incompressible liquid (possessing additionally the superfluidity) when the Dirac fundamental quantization scheme \cite{Dir} is applied to the to the Minkowskian  Higgs model.

To ground the latter statement, it will be useful to recall hydrodynamics.

In the monograph \cite{gidrodinamika}  (in \S 9) there was analysed the potential motion in a liquid. \par
It turned out that it is necessary and sufficient that ${\rm rot}~{\bf v}=0$ in the whole space (with ${\bf v}$ being the velocity of the liquid) in order for the motion of a liquid to be potential in the whole considered space. \par
 As for each  vector field possessing  the zero curl,  the velocity of a potentially moving liquid may be expressed \cite{gidrodinamika} as the gradient of a scalar: 
$${\bf v}={\rm grad}~ \phi. $$
On the  other hand, the Bianchi identity 
\be \label{Bianchi}
 D~ B=0
 \ee 
   for the (vacuum) "magnetic" tension $\bf B$ implies that  the Minkowskian BPS monopole vacuum suffered the Dirac fundamental quantization \cite{Dir} is an incompressible liquid\rm. \par
Really (see \S 10 in \cite{gidrodinamika}), the well-known  continuity equation
$$ \frac{\partial \rho }{\partial t}+ {\rm div} \rho {\bf v}=0$$
  (with $\rho $ being the density of the considered liquid) is simplified in a radical way as $\rho = {\rm const}$ (i.e. when the density remains constant along the whole volume besetting by the liquid during the whole time of motion). \par 
In  this case the continuity equation acquires the  simplest look 
\be
 \label{idealn}
 {\rm div}~ {\bf v}=0.
\ee 
As a consequence of  latter Eq., one gets  
\be
 \label{ideall1}
 \Delta \Phi=0
 \ee  
for a Higgs scalar field $\Phi$ (in particular, for that taking the shape of a vacuum BPS monopole).\par 
 Comparing  latter two Eqs. with Eqs. (\ref{Bianchi}) and (\ref{Gribov.eq}), 
 we see distinctly a parallel between the Minkowskian BPS monopole vacuum (suffered the Dirac fundamental quantization \cite{Dir}) and an incompressible liquid (possessing simultaneously manifest superfluid properties according the Gribov ambiguity equation (\ref{Gribov.eq})). \par
Meanwhile the Bogomol'nyi equation (\ref{Bog}) describes only superfluid properties of the Minkowskian BPS monopole vacuum. \par
\medskip
From the topological viewpoint, the property to be an incompressible liquid comes for the Minkowskian BPS monopole vacuum quantized by Dirac to the existence definite topological invariants in the appropriate Minkowskian  Higgs model. \par 
In a topologically nontrivial (non-Abelian)  gauge theory, such values always exist, invariant with respect to continuous deformations of (non-Abelian) fields. \par
The  one of most important topological invariants one encounters in a gauge theory is the { \it winding number functional}.\par
In Ref. \cite{Pervush2} it was defined as
\be \label{wind} X[A]=-\frac {1}{8\pi ^2}\int\limits _ {V} d^3 x \epsilon ^{ijk} {\rm tr}~ [{\hat A}_i \partial_j{\hat A}_k-  \frac {2}{3}{\hat A}_i{\hat A}_j {\hat A}_k],\quad A_{\rm in,~out}= A(t_{\rm in,~out},{\bf x}).\ee 
Herewith, without loss the generality, it may be set $ t_{\rm in,~out}\to\pm \infty$ for appropriate time instants. \par 
Knowing the winding number functional $ X[A]$, one can specify than the {\it Chern-Simons functional} (the {\it Pontryagin index} in the terminology \cite{Pervush2, Pervush3}) 
\be \label{Ch-S} \nu[A]=\frac {g^2}{16\pi ^2}\int\limits _{t_{\rm in}} ^{t_{\rm out}} dt\int \limits _ {V} d^3 x F_{\mu \nu}^a 
{\tilde F}^{\mu \nu}_a = X[A_{\rm out}]-X[A_{\rm in}]= n(t_{\rm out})-n(t_{\rm in}), \ee
involving the (YM) tension tensor $ F_{\mu \nu}^a $ and its dual, $ \tilde F_{\mu \nu}^a$.  

Alternatively to Eq. (\ref{wind}) for the winding number functional $ X[A]$, the (Pontryagin) degree of a map may be defined in a non-Abelian gauge theory as \cite{Al.S., Pervush2}
\be 
\label{degree}
{\cal N}[n]=- \frac {1}{24\pi ^2}\int_Vd^3 x \epsilon ^{ijk} ~{\rm tr}~ [L^n_iL^n_jL^n_k] \in {\bf Z}, 
\ee 
with
\be \label{cl.vac} {\hat A}_i \Rightarrow  L^n_i\equiv v^{(n)}({\bf x})\partial_i v^{(n)}({\bf x})^{-1} \quad {\rm as}~ \vert {\bf x}\vert \to\infty \ee
being the (classical) purely gauge vacuum configuration in this non-Abelian gauge theory. 

\medskip
There may demonstrated (see Refs. \cite {David2, Pervush2} and the article  by Jackiw "Topological investigations of quantized gauge theories" in \cite{Kurr}) that upon performing gauge  transformations (\ref{cl.vac}), the winding number functional $ X[A]$, (\ref {wind}), takes the  look 
\be
\label{chang} X[A^{(n)}_i]= X[A^{(0)}_i]+{\cal N}(n)+\frac {1}{8\pi ^2}\int d^3x \epsilon^{ijk}  {\rm tr}~ [\partial _i({\hat A}_j^{ (0)} L^n_k)]. \ee 
Let us denote, following \cite{David2}, as
\be 
\label{chang1}
 {\cal N}(n)+\frac {1}{8\pi ^2}\int d^3x \epsilon^{ijk}
{\rm tr}~  [\partial _i({\hat A}_j^{ (0)} L^n_k)]\equiv { N}({\cal N},A) 
\ee
two latter items in  (\ref{chang}). 

On the other  hand, according to
(\ref{wind}),  the winding number functional  $X[A^{(n)}_i]$ would take integers. 

Thus we come to the so-called \it self-consistency condition \rm \cite{David2}
\be \label{scc}
N({\cal N},A)=N, \quad N\in {\bf Z}. 
\ee
To satisfy (\ref{chang}),  we may choose the third item in  (\ref{chang}) to be \cite{David2} 
\be
  \label{sp}
N-\frac{\sin(2\pi N)}{2\pi};
 \ee
  then  the second item therein should be equal to  
\be
  \label{sp1}
 \frac{\sin(2\pi N)}{2\pi}. 
\ee
\medskip
In the Minkowskian Higgs model, implying always the spontaneous breakdown of the initial (for instance, $SU(2)$) gauge symmetry, one  encounters some more topological invariant: the magnetic charge $\bf m$.

Generally, it may be specified over a (Higgs-YM) field configuration $(\Phi,A)$ as \cite{Al.S.}
\be  
\label{mt1} 
{\bf m} (\Phi,A)= C~ \zeta (\Phi,A), \quad \zeta (\Phi,A)\in {\bf Z}; \ee
with $C$ being a constant that doesn't depend on the (Higgs-YM) field configuration $(\Phi,A)$.

Thus any magnetic charge $\bf m$ proves to be a topological invariant due to this independence of $C$ on $(\Phi,A)$, i.e. on fluctuations (deformations) of these fields \footnote{As a rule, one considers linear continuous deformations of fields: for example \cite{Al.S.}, 
$$ A_\mu^a({\bf x})\to t\tilde A_\mu^a({\bf x})+(1-t) A_\mu^{a'}({\bf x}),         $$
with $t\in [0,1]$.}.

Indeed, latter Eq. is not obvious, and we recommend our readers the monograph \cite{Al.S.} where it was derived in \S$\Phi$7.

In particular, in the Minkowskian Higgs model involving vacuum BPS monopole 
solutions, the magnetic charge $\bf m$ determines the Bogomol'myi lowest bound (\ref{Emin}) of the YMH energy. 

If this model is suffered the Dirac fundamental quantization \cite{Dir}, of the YMH (vacuum) energy  is like the density of an incompressible liquid \cite{gidrodinamika}: herewith the properties of an incompressible (superfluid) liquid come, in the Minkowskian YM theory involving vacuum BPS monopoles, to the properties of topological (deformation) invariants: magnetic charges $\bf m$, (\ref {mt1}), degrees of the map \cite{rem1,LP1}
\be
\label{top3}
\pi_2 S^2= \pi_3 (SU(2))=\pi_1(U(1))=\pi_1~S^1=\bf Z, 
\ee
referring to the $U(1)\subset SU(2)$ imbedding, and the winding number functional  (\ref {wind}).  
\par  
Of course,  enumerated topological invariants are present in each Minkowskian  Higgs model involving BPS monopole modes, but the incompressibility property may remain paid no heed in these models  considered in the heuristic quantization scheme \cite{FP1}.
\section{Properties of Gribov topological multipliers.}
There may shown \cite {LP2,LP1,David2,David3,Pervush2} that  the Gribov ambiguity equation (\ref{Gribov.eq}) 
together with the topological condition
\be
\label{X[n]}
X[A ^{(n)}_k]=n; \quad n\in {\bf Z};
\ee
are compatible with the unique solution \footnote{ Indeed, it is unique to within the whole family of gauge fields cohomological to each other \cite {Al.S.,LP1}.

This  family consists of such gauge fields that two 1-forms $\omega_1\equiv A_\mu d^\mu$ and $\omega_1'\equiv A'_\mu d^\mu$, involving  (transverse) YM fields $A_\mu$ and $ A'_\mu$, respectively, belonging to a one class of cohomologies differ on the exact 1-form $d\sigma$: $\omega_1-\omega'_1=d\sigma$ \cite {Al.S.}.

Latter Eq. may be rewritten approximately as
$$\partial_\mu (\omega_1-\omega'_1)= \partial_\mu d\sigma=0,$$ 
since $d\cdot d \sigma =0$ due do the Poincare lemma \cite {Al.S.}.

The analysis of the cohomological structure of gauge fields in the Minkowskian Higgs model quantized by Dirac was started in the paper \cite {LP1}.

The author intends to continue this analysis in a one of the next works. 
} to the classical YM equations of motion. 

The nontrivial solution to the equation for the Gribov phase ${\hat \Phi}_0(r)$ is well-known in this case \cite {LP2,LP1,Pervush2}:
\be
 \label{phasis}
{\hat \Phi}_0(r)= -i\pi \frac {\tau ^a x_a}{r}f_{01}^{BPS}(r), \quad 
f_{01}^{BPS}(r)=[\frac{1}{\tanh (r/\epsilon)}-\frac{\epsilon}{r} ].
 \ee
 It is just a $U(1)\subset SU(2)$ isoscalar "made" of  Higgs vacuum BPS monopole modes.

As a definite linear combination of these vacuum BPS monopole modes, the Gribov phase (\ref{phasis}) satisfies actually the Gribov ambiguity equation (\ref{Gribov.eq}).
\par
In Ref. \cite {David3} there was shown  that the function $ f_{01}^{BPS}(r)$, entering Eq. (\ref{phasis}) for the Gribov phase ${\hat \Phi}_0(r)$, has the asymptotic  
\be \label{bcf0}
f_{01}^{BPS}(0)=0; \quad f_{01}^{BPS}(\infty)=1.
\ee
\medskip
In the series of papers (for, instance, \cite {Pervush1, Azimov, Pervush3,Arsen}), there was demonstrated that the Gribov exponential topological multipliers $ v^{(n)}({\bf x})$ (entering Dirac variables $A^D$) satisfy the boundary condition
\be
 \label{bondari}
 v^{(n)}({\bf x})\to \pm 1, \quad \vert {\bf x}\vert \to \infty.
\ee 
The one way to prove latter Eq. was pointed out in Ref. \cite {Pervush3}.

Ibid it was proposed the following representation for Gribov multipliers $ v^{(n)}({\bf x})$:
\be
 \label{Axieser}
v^{(n)}({\bf x}) = \cos (\pi n ~f_{01}^{BPS}(r)) - i n^a \tau_a \sin(\pi n ~f_{01}^
{BPS}(r)); \quad n^a=x^a/r. 
\ee
The "technology" deriving latter Eq. is enough simple.  It was applied, for instance, in the monograph \cite{A.I.}, in \S 7.1, at the analysis of $SU(2)$ (global) rotations of a spinor $\varphi$. 

As in that case, one would take account of the relations
$$   ({\bf n}\tau)^{2k}=1; \quad ({\bf n}\tau)^{2k+1}={\bf n}\tau; \quad k\in {\bf Z}  $$
and expand $\cos (x)$ and $\sin(x)$ in the series.

Then the boundary condition (\ref{bondari}) for Gribov exponential topological multipliers $ v^{(n)}({\bf x})$ follows immediately from the spatial asymptotic (\ref {bcf0}) \cite {David3} for $ f_{01}^{BPS}(r)$ (when $\vert {\bf x}\vert \to \infty$).

\medskip
The alternative way to demonstrate the spatial asymptotic (\ref{bondari}) for $ v^{(n)}({\bf x})$ is recasting them, following \cite{Azimov}, to the look
\be 
\label{newfact}
v^{(n)}({\bf x})= \exp(\hat \lambda~_ {n,\phi_i} ({\bf x})),
 \ee 
with
\be
 \label{ln}
\hat\lambda~_{n,\phi_i} ({\bf x})\equiv i \tau ^a \Omega _{ab}(\phi_i)  \frac{x^b}{r} f_{01}^{BPS}(r)~ \pi n
\ee 
and 
$$ (\tau ^a)^\alpha_\beta ~\Omega _{ab}(\phi_i)= (u(\phi_i))_\gamma ^\alpha  (\tau ^a)^\gamma _\delta (u^{-1}(\phi_i))_\beta ^\delta, $$  
\be
 \label{eiler}
 (u(\phi_i)) ^\alpha_\beta = (e^{i\tau_1 \phi_1 /2})_\gamma ^\alpha  (e^{i\tau_2 \phi_2 /2})^\gamma _\delta (e^{i\tau_3 \phi_3 /2})_\beta ^\delta.
 \ee 
Here $\phi_i$ ($i=1,2,3$) are three Euler angles fixing the position of the coordinate system in the $SU(2)$ group space \footnote{ Indeed, we should always remember that "large" matrices 
$v^{(n)}({\bf x})$ belong to the residual $U(1)$ symmetry group, embedded in the initial $SU(2)$ group. }.\par
  To achieve the necessary asymptotic (\ref{bondari}) for Gribov exponential topological multipliers $ v^{(n)}({\bf x})$ at the spatial infinity one would impose the appropriate conditions onto the  Gribov phase in Eq. (\ref{newfact}).

More precisely, such conditions may be imposed onto the Euler angles $\phi_i$:  Gribov topological factors $v^{(n)}({\bf x})$ become \rm $\pm 1 $ at the spatial infinity when \rm  
\be
 \label{fas.ysl-e}
 \tau_i \phi_i = 4\pi n \quad ({\rm respectively,}\quad \tau_i \phi_i =2\pi n); \quad n\in {\bf Z}  \ee
\medskip

Indeed, one would take account of the fundamental constants $\hbar$ and $c$ at writing down  Gribov {\it exponential} topological multipliers $ v^{(n)}({\bf x})$, would be dimensionless in their definition.

In Ref. \cite{Pervush3} it was shown how to take account of $\hbar$ and $c$ in Gribov  topological multipliers $ v^{(n)}({\bf x})$, (\ref{degeneration1}), and now we shall stick to the arguments \cite{Pervush3} at statement the problem.

First of all, the mentioned fundamental constants enter various strong interaction models (YM and QCD) as $g/(\hbar c)$ in the lowest order of the perturbation theory. 

The said allows redefine gauge fields $A$ in terms of the strong interaction coupling constant $g/(\hbar c)$, as it was done, for instance, in (\ref {hatka}) \cite{Pervush3}.

That is why it is quite reasonable to recast Gribov { exponential} topological multipliers $ v^{(n)}({\bf x})$, (\ref{degeneration1}), in such a wise  that the dimension of the Gribov phase $\hat\Phi _0({\bf x})$, (\ref{phasis}), compensates the dimension of the coupling constant $g/(\hbar c)$, will now enter Gribov { exponential} topological multipliers $ v^{(n)}({\bf x})$: thus these remain dimensionless upon recasting.

On the other hand, such look of   Gribov topological multipliers  $ v^{(n)}({\bf x})$ ensures necessary properties of topological Dirac variables  (\ref{degeneration}), (\ref{degeneration1}): in particular,  that they are manifestly transverse and gauge invariant.

Additionally, the spatial asymptotic (\ref{bondari}) for $ v^{(n)}({\bf x})$ would be taken into account at this recasting.

Following \cite{Pervush3}, we write
\be \label{quantera}  v^{(n)}({\bf x})=\exp [n\hat \Phi _0({\bf x}) g/(\hbar c)].
\ee
It is easy to see \cite{Pervush3} that  the Gribov phase $\hat \Phi_0({\bf x})$, given by (\ref{phasis}),  may be recast to the look $2 \tau^a \lambda_a$, with 
\be \label{la}
\lambda_a = -i(\pi/2) ~n_a f_{01}^{BPS} (r).   
\ee
In this case, it is necessary to multiply $\lambda_a $ by $\hbar c/g$ (the value got in this way we shall denote as $\lambda'_a\equiv (\lambda_a \hbar c)/g$) and substitute then in (\ref{quantera}) in order for Gribov topological multipliers $ v^{(n)}({\bf x})$, given through (\ref{quantera}), to be, indeed, dimensionless.

By  that we come back to Eq.  (\ref{phasis}) for the Gribov phase $\hat \Phi_0({\bf x})$, that is free from $g$, $\hbar$, $c$. Respectively,  Gribov topological multipliers $ v^{(n)}({\bf x})$ become dimensionless.

Note that the value \cite{Pervush3}
\be \label{hatl}
\hat \lambda =  \lambda'_a \tau^a \frac{ g}{\hbar c} 
\ee
may be read from (\ref{quantera}),  (\ref {la}).

In this case Gribov exponential multipliers $ v^{(n)}({\bf x})$ acquire the look \cite{Pervush3}
\be \label{quantera1}
v^{(n)}({\bf x})=\exp(2n \hat \lambda); \quad n\in {\bf Z}; \ee
in terms $\hat \lambda$ (free from $\hbar$ and $c$, as it should be indeed).

The said  is in a good agreement with the definition (\ref{hatka}) \cite{Pervush3} of gauge fields $\hat A$ and with the property to be gauge invariant and transverse for the topological Dirac variables (\ref{degeneration}). 
\section{Discussion.}
Finishing  this study, we should like tell our readers about our further plans in developing the fundamental quantization formalism  \cite{Dir} for the Minkowskian Higgs model with vacuum BPS monopoles.

The outlines of this fundamental quantization were stated recently in Ref. \cite{fund}, and our readers can  obtain the general  idea, at reading this paper, how to quantize the Minkowskian Higgs model with vacuum BPS monopoles and which consequences implies such fundamental quantization.

 Now we shall narrate about our nearest investigations in this direction.

They will concern studying the nontrivial topological dynamics inherent in the \linebreak Minkowskian Higgs model with vacuum BPS monopoles quantized by Dirac \cite{Dir}. 

Herewith the arguments \cite{LP2,LP1, fund, David2, David3, Pervush2} will be repeated and extended.

It   will be argued that the origin of the mentioned nontrivial topological dynamics proves to be in resolving the YM Gauss law constraint (\ref{Gauss}) in the covariant Coulomb gauge (\ref{Aparallel}), (\ref {transv}).

In particular, topological Dirac variables $A^D$, (\ref{degeneration}), satisfy the Coulomb gauge (\ref{Aparallel}), (\ref {transv}).  

Thus resolving the YM Gauss law constraint (\ref{Gauss}) in terms of topological Dirac variables (\ref{degeneration})
  turns the YM Gauss law constraint (\ref{Gauss}) into the second-order homogeneous differential equation 
$$   [D^2_i(\Phi ^{(0)})]^{ac} A_{0c}=0,        $$
permitting the family of so-called {\it zero mode   solutions} \cite{Pervush2, Pervush1} 
\be\label{A0} A_0^c(t,{\bf x})= {\dot N}(t) \Phi_0^c ({\bf x})\equiv Z^c,
  \ee
implicating the topological variable $\dot N(t)$ and Higgs (topologically trivial) vacuum Higgs BPS monopole modes $\Phi_0^a ({\bf x})$.

$A_0$, specified in such a wise, may be treated as temporal components of gauge fields additional to  those equal to zero \cite{David2},  (\ref{udalenie}), got course the Dirac removal.

It is also the merit of the Dirac fundamental quantization  \cite{Dir} of the Minkowskian Higgs model (with vacuum BPS monopole solutions).

Recall in this context \cite{fund} that at the FP "heuristic" quantization \cite{FP1}
 of  Minkowskian Higgs models with monopoles, $A_0$ components of gauge fields are ruled out via fixing the Weyl gauge $A_0=0$.
\medskip

YM potentials $A_0$, (\ref{A0}), referring actually to the BPS monopole vacuum, induce specific $F_{i0}^a$ components of the YM tension tensor, taking the shape of so-called vacuum "electric" monopoles \cite{LP2,LP1}
$$   F^a_{i0}={\dot N}(t)D ^{ac}_i(\Phi_k ^{(0)})\Phi_{0c}({\bf x}).         $$
Issuing from vacuum "electric" monopoles $F_{i0}^a$, one can construct \cite{LP2,LP1, fund, David2}
the action functional
$$W_N=\int d^4x \frac {1}{2}(F_{0i}^c)^2 =\int dt\frac {{\dot N}^2 I}{2},$$
involving the  rotary momentum \cite{David2}
$$ I=\int \sb {V} d^3x (D_i^{ac}(\Phi_k^0)\Phi_{0c})^2 =
\frac {4\pi^2\epsilon}{ \alpha _s}
=\frac {4\pi^2}{\alpha _s^2}\frac {1}{ V<B^2>}.    $$
The YM coupling constant 
$$\alpha _s=\frac{g^2}{4\pi (\hbar c)^2 } $$
enters this expression for $I$.

The action functional $ W_N $ describes correctly collective solid rotations of the BPS monopole vacuum (suffered the Dirac fundamental quantization \cite{Dir}) with angular velocities ${\dot N}(t)$, proves to be constant (as the author intend to demonstrate).

In demonstrating constancy of $ {\dot N}(t)$, the arguments \cite{Pervush3} will be repeated.

The principal result will be got that \cite{Pervush3}
\be \label{dotn} {\dot N}(t)={\rm const} = (n_{\rm out}-n_{\rm in})/T\equiv \nu/T  \ee
where $n_{\rm out}$, $n_{\rm in}$~ $\in {\bf Z}$ refer to the fixed time instants $t=\pm T/2$, respectively.

It will be argued herewith (due to general QFT reasoning) that it would be set $T\to\infty$.

This implies  actual approaching zero by angular velocities ${\dot N}(t)$ for collective solid rotations inside the BPS monopole vacuum. 

\medskip
The crucial point in grounding Eq. (\ref{dotn}) is investigating the explicit look of  ${ N}(t)$, the noninteger degree of the map referring to the $U(1)\subset SU(2)$ embedding \cite{Pervush3}:
\bea
\label{winding num.}
\nu[A_0,\Phi^{(0)}]&=&\frac{g^2}{16\pi^2}\int\limits_{t_{\rm in} }^{t_{\rm out} }  dt  \int d^3x F^a_{\mu\nu} \widetilde{F}^{a\mu \nu}=\frac{\alpha_s}{2\pi}
 \int d^3x F^b_{i0}B_i^b(\Phi^{(0)})[N(t_{\rm out}) -N(t_{\rm in})]\nonumber \\  &&
 =N(t_{\rm out}) -N(t_{\rm in}). \eea 
\medskip 

The topological variable $N(t)$, specified via (\ref{winding num.}), determines the purely real (i.e. unambiguous {\it physical}) energy-momentum spectrum of the free rotator $W_N$:
$$ P_N ={\dot N} I= 2\pi k +\theta; \quad \theta  \in [-\pi,\pi];      $$
accompanied by the wave function
$$   \Psi _N=\exp (iP_N N).            $$ 
\medskip
The general origin of collective solid rotations inside the BPS monopole vacuum suffered the Dirac fundamental quantization \cite{Dir} is in the {\it Josephson effect} \cite{Pervush3}, coming to persistent circular motions of material points (quantum fields may be considered as a particular case of such material points) without (outward) sources. \par
These persistent circular motions of  quantum fields are characterized \cite{fund,Pervush3} by never vanishing (until $\theta  \neq 0$) momenta 
$$ P=   \hbar ~\frac {2\pi k+ \theta} L, $$
with $L$ being the length of the whole closed line along which the given quantum field moves. \par
Such momenta $P$ attain their nonzero minima 
$ p=\hbar \theta /L $ 
as $k=0$ and if $\theta  \neq 0$. \par
\medskip
The investigations about the Josephson effect are also planed for the future. \par
Repeating the arguments \cite{Pervush3}, it will be shown \cite{David2} that besides collective solid rotations inside the BPS monopole vacuum suffered the Dirac fundamental quantization \cite{Dir} (these rotations are characterized constant angular velocities $\dot N(t)$), the Josephson effect in the appropriate Minkowskian Higgs model comes to never vanishing (until $\theta  \neq 0$) vacuum "electric" fields ("electric" monopoles)
$$   (E_i^a)_{\rm min}= \theta \frac {\alpha_s}{4\pi^2\epsilon} B_i^a; \quad -\pi\leq \theta \leq \pi.
      $$
Such minimum value of the vacuum "electric" field $\bf E$ corresponds to trivial topologies $k=0$, while generally \cite{David2},
$$ F^a_{i0}\equiv E_i^a=\dot N(t) ~(D_i (\Phi_k^{(0)})~ \Phi_{(0)})^a= P_N \frac {\alpha_s}{4\pi^2\epsilon} B_i^a (\Phi _{(0)})= (2\pi k +\theta) \frac {\alpha_s}{4\pi^2\epsilon} B_i^a(\Phi_{(0)}).
       $$
\section*{Acknowledges.}
I am  very grateful to Profs. V. N. Pervushin and V. I. Tkach for the fruitful discussion and the series of useful remarks about my recent publications. 

\end {document}